\documentclass[journal]{vgtc}                     % final (journal style)
\onlineid{0}

\vgtccategory{Research}

\vgtcpapertype{research}

\title{Gridlines Mitigate Sine Illusion in Line Charts}

\author{%
  Clayton Knittel, 
    Jane Awuah, % \authororcid{Clayton Knittle}{0000-0002-1825-0097},
  Steven Franconeri, and 
  Cindy Xiong Bearfield
}

\authorfooter{
  \item
  	Cindy Xiong Bearfield and Jane Awuah are with the Georgia Institute of Technology.
  	E-mail: cxiong, jawuah3@gatech.edu
  \item
  	 Clayton Knittle is with Google.
  \item Steven Franconeri is with Northwestern University. 
}

\abstract{%
  The sine illusion is an underestimation of the difference between two lines when both lines have increasing slopes. 
  We evaluate three visual manipulations on mitigating sine illusions: dotted lines, aligned gridlines, and offset gridlines via a user study.
  We asked participants to compare the deltas between two lines at two time points and found aligned gridlines to be the most effective in mitigating sine illusions. 
  Using data from the user study, we produced a model that predicts the impact of the sine illusion in line charts by accounting for the ratio of the vertical distance between the two points of comparison.
  When the ratio is less than 50\%, participants begin to be influenced by the sine illusion. This effect can be significantly exacerbated when the difference between the two deltas falls under 30\%. 
  We compared two explanations for the sine illusion based on our data: either participants were mistakenly using the perpendicular distance between the two lines to make their comparison (the perpendicular explanation), or they incorrectly relied on the length of the line segment perpendicular to the angle bisector of the bottom and top lines (the equal triangle explanation).
  We found the equal triangle explanation to be the more predictive model explaining participant behaviors.
}


\teaser{
  \centering
  \includegraphics[width = 0.85\linewidth]{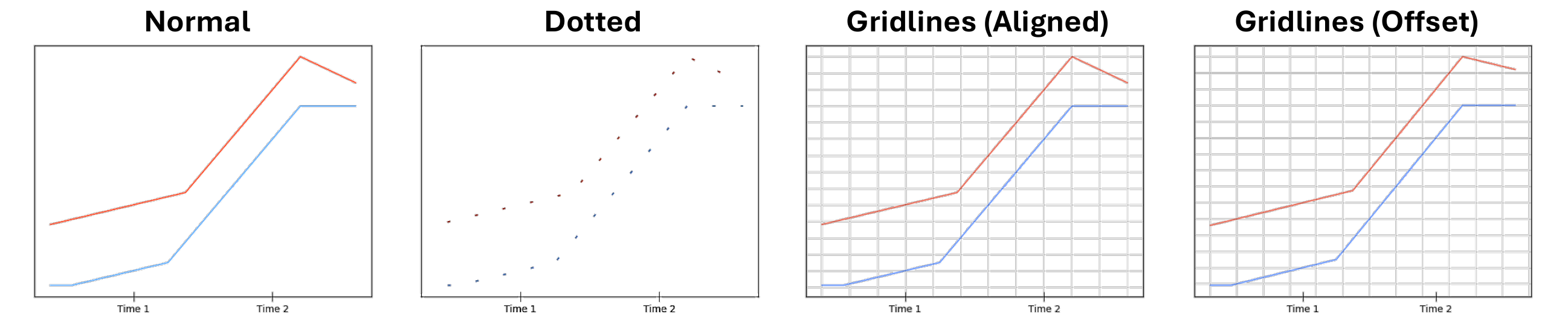}
  \caption{Four conditions: default, dotted, gridlines (aligned and offset) tested in this paper.}
  \alt{The default line chart shows a line chart with two lines, one on top and on on the bottom. Both lines folds into two segments, such that the second segment is steeper than the first. The dotted line chart shows the two lines as dots spaced out along the line without lines connecting them. The gridline charts are the default line charts with vertical and horizontal guidelines behind them. The gridelines can be aligned or offset. Aligned guidelines have vertical gridlines fall right on the comparison points, Time 1 and Time 2, where as the vertical gridlines in the offset guidelines condition do not fall right on the two comparison points.}
 \label{fig:All graphs}
}

\graphicspath{{figs/}{figures/}{pictures/}{images/}{./}} % where to search for the images

\usepackage{tabu}                      % only used for the table example
\usepackage{booktabs}                  % only used for the table example
\usepackage{lipsum}                    % used to generate placeholder text
\usepackage{mwe}                       % used to generate placeholder figures
\usepackage{accessibility}
\usepackage{mathptmx}                  % use matching math font
\newcommand{\add}[1]{{\textcolor{black}{#1}}}

\begin{document}

%%%%%%%%%%%%%%%%%%%%%%%%%%%%%%%%%%%%%%%%%%%%%%%%%%%%%%%%%%%%%%%%
%%%%%%%%%%%%%%%%%%%%%% START OF THE PAPER %%%%%%%%%%%%%%%%%%%%%%
%%%%%%%%%%%%%%%%%%%%%%%%%%%%%%%%%%%%%%%%%%%%%%%%%%%%%%%%%%%%%%%%

\firstsection{Introduction}

\maketitle

First formally introduced by Cleveland and McGill in 1984~\cite{cleveland1984graphical}, the sine illusion describes a perceptual error where more quickly changing pairs of lines can lead to bigger underestimates of the delta between them.
Researchers printed out vertical lines of equal length following a sine wave pattern~\cite{day1991sine} and noticed that the lines at the straight oblique sections appeared shorter than those in the turns. 
This illusion is common in the real world.
Multi-line line charts, stream graphs, and area charts all harbor opportunities for this bias, whenever viewers compare vertical distances between lines~\cite{vanderplas2015signs}.

% is often noted as a difficulty to visually compare the difference between two lines rather than a visual illusion 
% where pairs of lines with different rates of slope change plotted together lead to an underestimation of the delta between them. 

%happens when  where 

% first description of the basic effect to Cleveland & McGill (1984)~\cite{cleveland1984graphical}
% . Inspect the figure on the ←left (part of their Fig. 26): the two curves seem to get closer to each other going up – however, their vertical separation stays constant, the difference between the two is always the same. [Their top right area would correspond to the zero crossing area in the sine illusion above.] Cleveland & McGill write

% it's a common error in the real-world.
% has been frequently noted, though usually
% not as an optical illusion. Rather, the problem is typically identified as the difficulty of
% visually subtracting two curves (see e.g. ~\cite{vanderplas2015signs}, p. 35 or Cleveland and McGill 1984,
% p. 549), and the resulting erroneous conclusions when this process goes awry. Playfair’s
% chart of the balance of trade between England and the East Indies (Playfair and Corry,
% 1786; Playfair et al., 2005) (shown in Appendix A) represents the possibly oldest example of
% this common phenomenon. In more modern visualizations, bivariate area charts and “stream
% graphs” (Byron and Wattenberg, 2008) commonly produce the illusion (see an example at
% http://bl.ocks.org/mbostock/3894205).

When comparing the deltas, the sine illusion happens as viewers rely on one of many potentially irrelevant visual cues as proxies for the actual distance between two lines
% There are several possible explanations of `non-relevant' visual cues in this context.
For example, viewers might compare the areas of the regions near the two points of comparisons between the two lines rather than just the vertical distances, similar to the Müller-Lyer line illusion~\cite{day1991sine, howe2005muller}, where viewers include the arrow tips in the length comparison to overestimate the line length, and the `hull area' proxy presented by Jardine et al.~\cite{jardine2019perceptual}, where the viewer perceives an implied hull bounded by the bars when comparing the means of bar sets.
% When comparing the deltas, the sine illusion happens as viewers rely on one of many potentially irrelevant visual cues as proxies for the actual distance between two lines
% % There are several possible explanations of `non-relevant' visual cues in this context.
% For example, Day et al.~\cite{day1991sine} posit that the sine illusion can be explained as a perceptual comprise of the greater overall distance between the two lines around the points of comparison, similar to drawing out a circle around each point of comparison. 
% One suggestion is that viewers rely on surrounding the points of comparison similar to viewing a Müller-Lyer line illusion~\cite{day1991sine, howe2005muller} and the `hull area' proxy presented by Jardine et al.~\cite{jardine2019perceptual}. 
Others suggest that, instead of relying on the vertical distance between two points with a common X-axis value, viewers might rely on the orthogonal distance (or the minimal distance) which lead to non-veritical comparisons between points on the two lines that do not share a common x-axis value~\cite{vanderplas2015signs, bu2020sinestream}, similar to doing a Deming regression~\cite{reimann2022lollipops}.
We contrast these two explanations by pitting two models in competition to best model the sine illusion.
We refer to the two models as the `perpendicular model' and the `equal triangle' model, which we describe more closely in Section~\ref{twomodels}. 

% \noindent \textbf{(1) Perpendicular:} this heuristic is about taking the perpendicular distance between the two lines, anchored on the bottom line at Time 1 and at Time 2, as shown in the middle panel in Figure~\ref{fig:fhmodeling}. This idea is inspired by existing work that suggests that sine illusion happens because people rely on the orthogonal instead of the vertical distance between the two lines.
% \vspace{1mm}

% \noindent \textbf{(2) Equal Triangle:} this heuristic is about taking the length of the line segment perpendicular to the angle bisector of the bottom and top lines, as shown in the right-most panel in Figure~\ref{fig:fhmodeling}. This idea is inspired by the `hull area' proxy from Jardine et al.~\cite{jardine2019perceptual} and the theory proposed by Day et al.~\cite{day1991sine} which suggest that participants could have considered the overall general dimensional area surrounding the points of comparison when making the decision. 

Existing work has investigated visualization solutions to mitigate sine illusions.
% , rom the visualization perspective, we can design visualization tools to mitigate sine illusions
% it's critical because...
For example, Bu et al. developed Sinestream to reduce the effect of sine illusions in stream graphs~\cite{bu2020sinestream}. It manipulates the geometry of a stream graph by the bottom-most curve such that the orthogonal and vertical orientations of the lines align. 
In this paper, we test two more alternative designs to mitigate sine illusion: dotted lines and gridlines.
In the dotted lines design, we break the area surrounding the points of comparison by separating the lines into spaced dots. 
This increases the perceptual difficulty of viewers relying on overall dimensions (i.e., the `hull area' proxy) and areas to make their judgment.  
In the gridlines design, we add gridlines to the line chart such that the vertical lines can anchor and nudge viewers to compare the vertical distance rather than orthogonal distances between the two lines.
We also manipulate the ratio of vertical distance between the two lines at the two points of comparison.
This allows us to identify the threshold for when sine illusion begins to significantly interfere with a viewer's ability to correctly compare the deltas between two lines. 

\vspace{2mm}

\noindent \textbf{Contribution:} \add{We contribute an experiment demonstrating the sine illusion in line charts and model the severity of the illusion as a function of the ratio of the vertical distance between the two points of comparison. 
This work provides a perceptual foundation to inform future designs of visualization tools that mitigate sine illusions, as well as a quantitative model describing the influence of sine illusions as a function of the deltas between the two lines at the points of comparison.}

\begin{figure*}[ht]
    \centering
    \includegraphics[width = 0.9\linewidth]{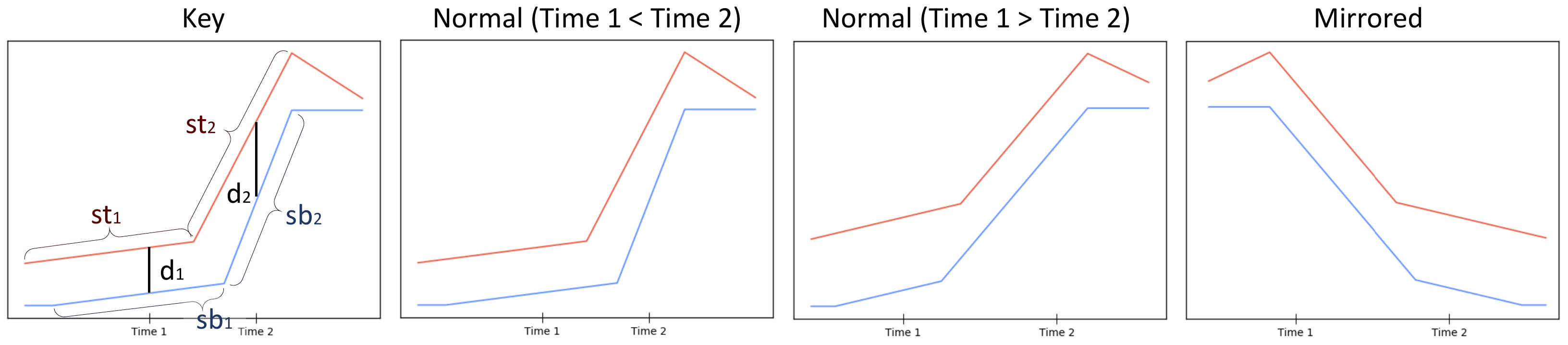}
    \caption{Left: Key explaining the line aspects we manipulated in this paper. Middle: Two examples showing conditions where the delta at Time 1 and Time 2 is greater respectively. Right: The mirrored condition of the Time 1 > Time 2 example.}
    \alt{The line charts in our experiment shows two lines, one on top, and on on the tobbom. Each line has two relevant segments. The top and the bottom lines have the two segments vertically assigned. We manipulated the line slopes such that the top line has slopes st1 and st2, where st1 < st2. The bottom line has slopes sb1 and sb2, where sb1 < sb2. We asked participants to compare the distance between the two lines at time 1, which is encompassed in the first line segment (with slopes st1 on top and sb1 on the bottom), and at time 2, which is encompassed in the second line segment (with slopes st2 on top and sb2 on the bottom). We refer to the distance at time 1 as d1 and the distance at time 2 as d2. The sine illusion is the illusion that d1 is greater than d2 when in reality the reverse is true. We also created a mirror image where we flipped the stimuli horizontally, such that line segment 1 goes on the right, and line segment 2 goes on the left. In the mirrored condition, d2 appears greater than d1 when the reverse is true.}
    \label{fig:key}
\end{figure*}

%-------------------------------------------------------------------------
%-------------------------------------------------------------------------
\section{Background and Related Work}

% \subsection{Perceptual Biases in Visualizations}
% - mention work from the human perception side and discuss illusions and heuristics people use to make decisions visually
% - might discuss visual attention mechanisms that contribute to visual illusions

% - more specific examples from data visualizations
% - relevance of visual illusions to visual analytics

% Data visualization is essential in communicating data, irrespective of size \cite{mccoleman2020no}. 
% While the type of data and the systematic bias of the designer determine the kind of visualization used to present it, communication is completed when the receiver correctly understands the illustration and interprets the data or results \cite{hong2021weighted}.
People can be cognitively and perceptually biased when interpreting data visualizations~\cite{valdez2017framework, wall2017warning, dimara2018task, xiong2022investigating, healey2011attention, quadri2021survey}.
For example, visualization readers can overly focus on salient features when making sense of data~\cite{healey2011attention}.
They can gravitate toward specific trends~\cite{bearfield2024same}, colors and highlights \cite{ajani2021declutter}, larger font sizes \cite{hegarty2010thinking}, and specific annotations~\cite{stokes2023role} that are aligned with their beliefs and agendas \cite{xiong2019curse, xiong2022seeing}. %, giving in to letting their intuition drive decision making \cite{padilla2018decision, kahneman2011thinking}.

People are biased by their mental prototypes when making sense of data. %Perception of visualized data can be distorted by category repulsion.
For example, when participants are tasked with estimating the height of bars, they underestimate bars that are taller than they are wide and overestimate bars that are wider than they are tall, which suggests that they see a bar mark more as a prototypical square ~\cite{ceja2020truth}.
This effect generalizes to stacked bars and dot plots.
When participants were asked to visually redraw the y-position of stacked bars and dot plots, smaller values were overestimated and larger values were underestimated, with an average error of approximately 10\% ~\cite{mccoleman2020no}. 
When they were asked to reproduce the values verbally instead of visually, participants can be anchored by round numbers such as five's and ten's~\cite {savalia2022data}.
When reading line charts or regressions, people also tend to see the lines as closer to 45 degrees (often referred to as `bank to 45 degrees'~\cite{cleveland1993model}), as they see the lines as closer to a prototypical angle bisector~\cite{heer2006multi}.

These works on human perception of visualizations uncover the biasing effects of cognitive prototypes and provide insights for crafting more effective visualizations that mitigate biases. %the impact of these biases This study uncovered the cognitive processes underlying data reproduction and provided insights for crafting more effective visualizations that mitigate the impact of these biases 
For example, Heer et al.~\cite{heer2006multi} developed optimization techniques to reduce the banking to 45-degree bias by automatically identifying trends in data and generating a tailored chart scale. % that `banks' the data to 45 degrees. 
They migrated such a bias by aligning what people intuitively see in data with the objectively correct interpretation. 
% First, we propose alternate optimization criteria designed to further improve the visual
% perception of line segment orientations. Second, we develop multi-scale banking, a technique that combines spectral analysis
% with banking to 45°. Our technique automatically identifies trends at various frequency scales and then generates a banked chart
% for each of these scales. We demonstrate the utility of our techniques in a range of visualization
We take a similar approach in this paper to study sine illusions.

\add{The sine illusion is a perceptual distortion when viewers misjudge the alignment and spacing between a pair of data value, both of which follow a sine-wave pattern~\cite{vanderplas2015signs}.
This illusion could appear in non-sinusoidal curves as well, making it more omnipresent~\cite{hofmann2013common, cleveland1984, schonlau2024hammock}.
As shown in Figure~\ref{fig:key}, when comparing the distance at Time 1 (d1) and Time 2 (d2), the sine illusion can happen when the slopes of the lines increase over time (e.g.,$st_{1}$ < $st_{2}$, $sb_{1}$ < $sb_{2}$).  % ($st_{1}$ < $sb_{1}$, $st_{2}$ < $sb_{2}$, $st_{1}$ < $st_{2}$, $sb_{1}$ < $sb_{2}$).
This difference in slopes distorts the perceived distance between the two lines (commonly referred to as the `detla'). 
The delta at an earlier point is perceived to be larger than the delta at a later point, despite the reverse being true.
There has been some research aimed at mitigating its distortive effects.
For example, Reimann et al. demonstrated that visual aids like "lollipops" in scatterplots could align visual and statistical fit estimates, potentially reducing the impact of the sine illusion. ~\cite{reimann2022lollipops}
We join existing effort to study the perceptual phenomenon of sine illusions in visualizations, offering concrete guidelines to inform visualizations design that can mitigate this bias.}

\section{Experiment}
We conducted an experiment comparing the effectiveness of three visualization designs in mitigating the sine illusion in line charts with two lines.
We manipulated the ratio of the deltas between the lines at two points of the comparison. 
This allows us to obtain a threshold above which the sine illusion begins to take a significant effect in biasing viewer perception.
To better understand the underlying drivers of the sine illusion, we further model and compare two potential explanations of participants' behavior. 

\subsection{Participant}
We recruited 62 participants, with an average age of 20.65 ($SD$ = 4.98), for this study, using the online crowdsourcing platform Amazon's Mechanical Turk~\cite{mturk}. %<- need to figure out citation
Participants were compensated at a rate of 12 USD per hour. % for a XXX-minute survey. 
Thirty-one participants completed the study and contributed 50,777 trials of valid responses.

% \begin{figure}[htp]
%     \centering
%     \includegraphics[width=4cm]{figs/Demo1.png}
%     \caption{Graph illustrating line illusion}
%     \label{fig:key}
% \end{figure}

\subsection{Experimental Design}
We generated charts depicting two increasing lines (blue and red). 
% The red line on top has 3 segments while the blue line on the bottom has 4 segments. % and was at the bottom of the graph.
To establish the illusion task, we identified two time points on the chart (Time 1 and Time 2), and manipulated the vertical distance between the blue and the red line at these two points, as shown in Figure~\ref{fig:key}.
Participants were tasked with comparing the distance between the red and blue lines at both time points to determine whether the differences between them is larger at Time 1 or Time 2. 
We refer to the distance between the two lines at Time 1 to be $d1$, and the distance between the two lines at Time 2 to be $d2$.

To determine a threshold past which sine illusion begins to have a strong effect based on the ratio of $d1$ and $d2$, we manipulate their ratio between 0.5 to 1.0, based on pilot studies that suggest that a ratio below 0.5 between $d1$ and $d2$ leads to generally highly accurate performance. 
To avoid a combinational explosion of potential values to display for $d1$ and $d2$, we fixed the value of $d1$ to be 0.1 while varying the value of $d2$ at 0.025 increments, ranging from 0.075 to 0.2. 
This manipulation required us to change the slope of the second and third segments of the blue line, which we refer to as $sb1$ and $sb2$, as well as the first and second segments of the red line, which we refer to as $st1$ and $st2$, as shown in Figure~\ref{fig:key}.
To account for the potential effects of the slopes on participants' ability to compare the distances at Time 1 and Time 2 without the loss of generalizability, we randomly generated values of $sb1$ and $sb2$ to satisfy the following constraints: (1) the red line always remains above the blue line during Time 1 and Time 2, and (2) for each combination of $sb1$, $sb2$, $st1$, $st2$, three values remained constant while the fourth value is varied to allow us to control the specific effect of the varied component. 
\add{We created a charting space of 1.75 (x) by 2.25 (y) with 0.25 intervals.}
These two constraints produced $1638$ charts. 
% The values of $ma$ ended up ranging from 0.0513 to 0.9104.
% The values of $na$ ranged from 0.6305 to 4.0137.
% The values of $mp$ ranged from 0.0536 to 0.8942.
% The values of $m1$ ranged from 0.8829 to 2.8438.

With this current design, because both lines are positively slanted, their slopes at Time 1 are always smaller than their slopes at Time 2. \add{To account for potential left-right position-driven bias in participant response, we mirrored all of our charts so that both lines are negatively slanted in the mirrored conditions}, as shown in Figure~\ref{fig:key}.
This creates $1638*2=3276$ charts.
This counterbalancing also accounts for potential response biases to prevent participants from achieving a higher accuracy rate by only selecting Time 1. 

For each chart, we varied the design of the lines to be either default or dotted, or we added gridlines to the default version of the chart, as shown in Figure~\ref{fig:key}, creating $3276*3=9828$ charts.
For the $3276$ gridline charts, we recognize that whether the gridlines aligned with the marker at Time 1 and Time 2 also might have an effect on comparison accuracy.
Therefore, as a counterbalancing, for half of the gridline charts \add{(randomly selected from the stimuli pool)}, we aligned the vertical gridlines with Time 1 and Time 2, and for the other half, we offset them, as shown in Figure~\ref{fig:All graphs}.
\subsection{Procedure}
% In this experiment,  %filled out a survey with a similar structure. They received a link, which directed them to 
Participants were instructed to set their browser window to 100\% before proceeding into the study. 
We directed them to compare the revenue of two companies: A and B, over two time intervals, Time 1 and 2.
% ... (describe slide 2 and 3). 
% - make sure you describe the flow logic: if they got the answer wrong, what happens. If they got the answer right...
Participants were to ascertain the time at which the difference in revenue between A and B was bigger by clicking on the button for the selected time or pressing 'F' for Time 1 and 'J' for Time 2. 
Figure~\ref{fig:key} shows both a condition where the difference at Time 1 is bigger and a condition where the difference at Time 2 is bigger. 
% In the example in Fig. 3, the difference in revenue at Time 2 was bigger. 
All participants went through a practice trial with feedback.
% If they selected Time 2 as their answer when t, they would receive the pop-up message, ``Correct. The difference in revenue between A and B is Bigger in Time 2.'' 
% If they selected Time 1, they would receive the pop-up message ``Incorrect. The difference in revenue between A and B is bigger in Time 2.''

% The experiment starts after the instructions. 
Participants were also instructed to respond as accurately and fast as they could.
They could also take a break at any time from the experiment. 
% They were also instructed to click on the "Start Experiment " button, to begin the survey.
After reading the instructions, they proceeded to complete 1638 trials per person to keep the length of the experiment reasonably under one hour. 
% (look into R - the dataset to see how many trials per participant has completed).
After they had completed the experiment, they were given an MTruk completion code and were redirected to a Qualtrics survey to enter their demographic information.

\begin{figure}[ht]
    \centering
    \includegraphics[width = \columnwidth]{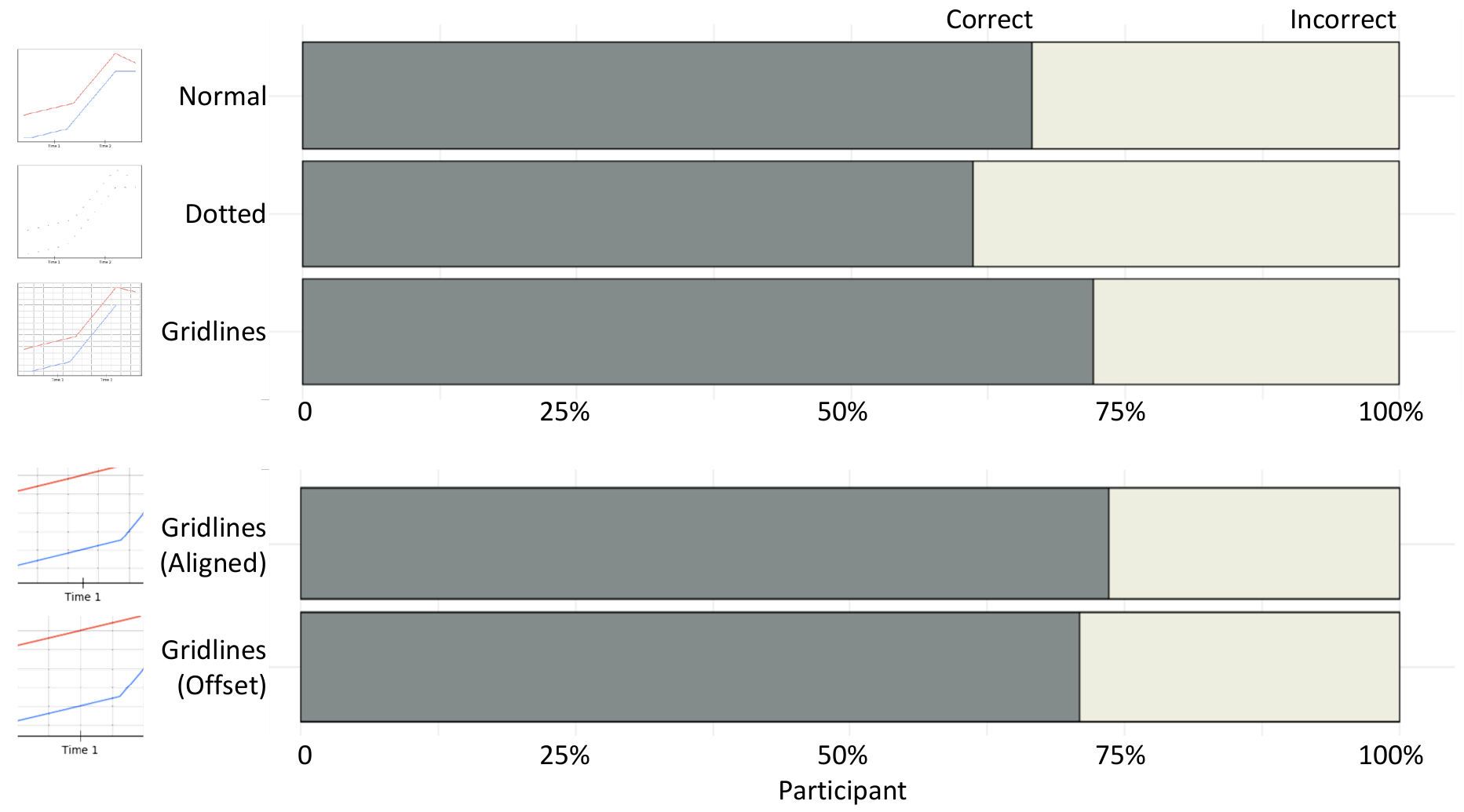}
    \caption{Overall accuracy for the three conditions and gridlines two counterbalancing conditions (aligned and offset).}
    \alt{Overall accuracy across all condition is above chance (which is 50\%). In general, participants performed the best in the gridlines condition, followed by the normal default condition, and the worst in the dotted condition. Amongst the two gridlines condition, participants performed slightly better with the aligned gridlines, with accuracy reaching about 75\%, compared to the offset gridlines condition.}
    \label{fig:conditionAccuracy}
\end{figure}

\begin{figure*}[ht]
    \centering
    \includegraphics[width = 0.8\linewidth]{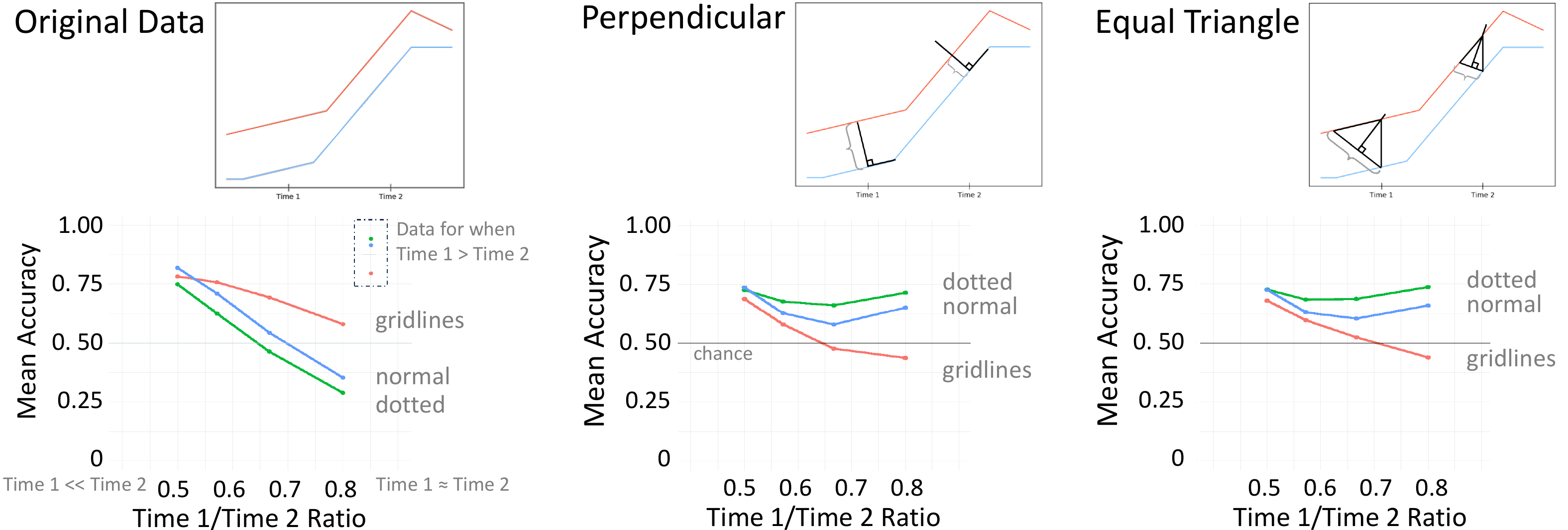}
    \caption{Accuracy by the ratio of deltas at Time 1 and Time 2, for the original data, in comparison to the two models: perpendicular and equal triangle. The higher performance accuracy under both models suggests that participants were relying on the orthogonal distance to make their comparisons.}
    \alt{We are comparing participants' response accuracy from the original data, to two modeled data: the perpendicular heuristic model, and the equal triangle heuristic model. The results are shown in three charts. All charts have the ratio of the deltas at Time 1 and Time 2 on the x-axis, which ranges from 0.4 to 0.9. Lower ratio means the actual delta at Time 1 is smaller than that at Time 2, and as the ratio approaches 1, the more similar the actual deltas are. The y-axis shows the mean accuracy, where 50\% is chance performance. For the original data, we see that mean accuracy linearly decrease with time 1 to time 2 ratio, dipping down to below chance when the ratio is between 0.6 and 0.7, for dotted and normal conditions. The gridline condition mitigates this bias, such that the performance stays comfortable above chance even when the ratio of time 1 to time 2 is greater than 0.8. If we model the accuracy by not the actual groundtruth accuracy, but by the perpendicular distance of the two lines, or by the equal triangle distance of the two lines, we see that the dotted and normal condition becomes associated with high performance, around 75\% across all values of time 1 to time 2 ratio. The gridlines performance, on the other hand, drops to below chance around 0.6 ratio between time 1 and time 2. This suggests that when people where doing the task with the dotted and normal charts, they are relying on the perpendicular and/or equal triangle heuristic rather than comparing the correct deltas. But when they have guidelines, they are able to rely less on these heurstics to compare the correct deltas.}
    \label{fig:fhmodeling}
\end{figure*}

%%%%%%%%%%%%%%%%%%%%%%%%%%%%%%%%%%%%%%%%%%%
\subsection{Results: Overall}
In analyzing our results, we filtered for trials that were completed in under 21791 milliseconds and more than 200 milliseconds as a quality control.
We picked the lower bound because this task of comparing Time 1 and Time 2 requires an eye movement, and existing work in human perception suggests that it takes about 200 mililiseconds for eye movement to begin~\cite{purves2001neuroscience}.
We picked the upper bound based on one standard deviation above the mean response time.
After filtering, we obtained 50,777 trials of valid responses, with an average response time of 2062.39 milliseconds. 

In our following analysis, we also filtered out the trials where the difference between the two lines at Time 1 equals that at Time 2.
These equality trials were added to prevent participants from consistently selecting one time over the other.
Since there is no valid correct answers in these equality trials, we excluded them from our analysis. 
In general, for these 7457 equality trials (14.7\% of the total trials), participants were 46.6\% likely to choose Time 1.

% filtering: 
% no NA on response, no NA no RT
% Reaction time:
% n = 50777 trials
% average rt = 2062.39
% sd rt = 19728.64	

% filter for RT < 200ms, upperlimit 1 SD above the mean = 21791.03, in this range
% The time course of a saccadic eye movement is shown in Figure 20.4. After the onset of a target for a saccade (in this example, the stimulus was the movement of an already fixated target), it takes about 200 ms for eye movement to begin~\cite{purves2001neuroscience}

% 1) Overall accuracy in terms of percentage for mirror and non-mirror
% accuracy rate for non-mirror = 4378/(4378+2025) = 0.6837
% accuracy rate for mirrored = 3974/(3974+2397) = 0.6238
% To compare whether these two accuracies are statistically significantly different, we conducted a t-test.
% The test reveals a difference such that participants responded with significantly higher accuracy in the non-mirrored condition ($t(12745) = 7.14$, $p < 0.001$). 

% 2) Does any of the slope values significantly impact accuracy?
% - compare accuracy per slope value 

% 3) Brainstorm some additional questions

As shown in the top section of Figure~\ref{fig:conditionAccuracy}, participants performed this task above chance (50\%) across all conditions. 
% accuracy (not mirrored)
% 0	2025			
% 1	4378	
% More correct than incorrect
In 66.5\% of the trials, participants drew correct conclusions from the normal condition, 61.2\% from the dotted-lines condition, and 72.1\% from the gridlines condition.
A logistic linear regression predicting accuracy with condition shows that, compared to the normal condition, participants are about 1.30 times more likely to obtain the correct answer with the gridlines condition (p $< 0.001$).
For the dotted-lines condition, they were only 0.79 times as likely to obtain the correct answer (p $< 0.001$).

\noindent \textbf{Takeaway:} Making lines dotted does not mitigate the sin illusion bias, but adding gridlines can help. 
% Gridelines help! 
% Dotted hurt

% compare 3 conditions' accuracy:
%   condition accuracy     n
%   <chr>        <dbl> <int>
% 1 dotted           0  5327
% 2 dotted           1  8386
% 3 gridlines        0  3296
% 4 gridlines        1  8522
% 5 normal           0  4235
% 6 normal           1  8406

% A chi-square test of independence with Bonferroni adjustment suggests that ...($\chi^2$ = 341.41, $p$ < 0.001). 
% varying proportion of participants drew correlation conclusions from different visualization designs 

% X-squared = 341.41, df = 2, p-value < 2.2e-16

%                    Odds ratio
% (Intercept)          1.984888
% conditiondotted      0.793115
% conditiongridlines   1.302622

% Overall, normal people are equally likely to get the answer correct and incorrect (residual $= 0.53$, p $> 0.05$).

% Logistic regression predicting accuracy with condition shows that compared to normal, 
% For gridlines, about 1.30 times more likely correct (p $< 0.001$).
% For dotted lines, about 0.79 times as likely correct (p $< 0.001$).
% dotted	Residuals	15.9776533	-15.9776533	
% dotted	p values	0.0000000	0.0000000	
% gridlines	Residuals	-16.0409898	16.0409898	
% gridlines	p values	0.0000000	0.0000000	
% normal	Residuals	-0.5302067	0.5302067	
% normal	p values	1.0000000	1.0000000	

%-------------------------------------------------------------------------
\subsection{Mirroring}
\add{We also mirrored all our displays to counterbalance our experiment accounting for the effect of the x-axis order (relative horizontal position of the sine illusion bias) for Time 1 and Time 2, as shown in Figure~\ref{fig:key}.}
% A chi-square test of independence with Bonferroni adjustment suggests that participants were more likely to get the answer correct than not in both the default and mirrored conditions
% ($\chi^2$ = 126.31, $p$ < 0.001). 
On average, in  68.8\% of the trials, participants obtained the correct answer from the default normal condition, and in 63.9\% of the trials, participants obtained the correct answer from the mirrored condition.
A logistic regression revealed that participants were only 0.78 times as likely to get the answer correct when the figure is mirrored, compared to the default (Est = $-0.244$, SE = $0.022$, p < $0.001$). 
This suggests that the x-axis ordering and relative line positions may have a biasing effect to exaggerate sine illusions. 
We further discuss potential research avenues to better understand such bias in Section~\ref{limitations}.

% Future work -> maybe y-axis has an effect, when the sine illusion is occurs to the left...
% yes = 0.78no.

% counterbalancing mirrored accuracy:
% <chr>
% accuracy
% <dbl>
% n
% <int>
% n	0	5915		
% n	1	13187		
% y	0	6943		
% y	1	12127	

% X-squared = 341.41, df = 2, p-value < 2.2e-16

% Welch Two Sample t-test
% data:  x = mirrored and y = not mirrored (default!, left appears wider, but right is actually wider)
% t = -5.7217, df = 12113, p-value = 1.08e-08
% alternative hypothesis: true difference in means is not equal to 0
% 95 percent confidence interval:
%  -0.06581212 -0.03222591
% sample estimates:
% mean of x mean of y 
% 0.6392008 0.6882198 

%-------------------------------------------------------------------------
%-------------------------------------------------------------------------
% \section{Experiment 2 Comparing Regular Lines to Dotted Lines}
% - dotted lines
% - also compare to mirror

% %-------------------------------------------------------------------------
% %-------------------------------------------------------------------------
% \section{Experiment 3 Comparing Regular Lines with Grid Lines}
% - Yes Off-set Grid (3a) to No Off-set Grid (3b)
% - off-set gridlines

\subsection{Effect of Gridlines On/Offset}
% special case of gridlines 
% Exp 1b for gridline, compare on/off accuracy:
% - Yes Off-set Grid (3a) to No Off-set Grid (3b)
% - off-set gridlines

We took a closer look at the gridlines condition to understand how it might have helped increase accuracy and mitigate the sine illusion.
As shown in Figure~\ref{fig:All graphs}, we manipulated the gridlines condition to either align the gridlines with Time 1 and Time 2 or offset it such that no vertical line goes through Time 1 or Time 2. 
We constructed a logistic regression predicting accuracy with whether the gridlines were offset or aligned. 
We found that aligning the gridlines increased accuracy to be 1.14 times that of the offset condition (p $= 0.00122$).
Overall, as shown in the bottom section of Figure~\ref{fig:conditionAccuracy}, participants performed with 73.53\% accuracy when the gridlines are aligned and with 70.86\% accuracy when the gridlines are offset.

% More likely to get it correct when offset = FALSE.
% x-offsetFALSE  = 1.142589 TRUE (odds ratio, p $= 0.00122$)
% 0.7353207\% accuracy when offset = FALSE
% 0.7085787\% accuracy when offset = TRUE
% x_offset
% FALSE	0	1465		
% FALSE	1	4070		
% TRUE	0	1831		
% TRUE	1	4452	

%-------------------------------------------------------------------------
\subsection{Relationship between Time 1 and Time 2 on Accuracy}
\label{twomodels}
We further investigate the driving factor behind the varying accuracy levels for the comparison task between Time 1 and Time 2.
Specifically, we examine how accuracy changes depending on the difference between the delta at Time 1 and the delta at Time 2. 
We created the stimuli by varying the ratio between Time 1 delta and Time 2 delta to range from 0.5 to 0.8.
For example, when the ratio between Time 1 delta and Time 2 delta is 0.5, it means the difference between the red line and the blue line at Time 1 is half that at Time 2. 
The closer this ratio is to 1, the more difficult this comparison task is.

As shown in the left-most panel in Figure~\ref{fig:fhmodeling}, as the ratio between Time 1 and Time 2 increases, the overall accuracy decreases for all three conditions.
However, the decrease is less steep for the gridlines condition compared to the normal and dotted conditions. 
% With one unit increase in the ratio between Time 1 and Time 2, the performance accuracy in the gridlines condition decreases to be around 3.37\% of the b. %0.03709983
One unit increase in the ratio between Time 1 delta and Time 2 (delta i.e., the difference between Time 1 and Time 2 becomes smaller) leads to the accuracy in task performance to be 3.37\% as accurate as the previous tier. 
For the dotted condition, task performance decreases to be only 1.40\% as accurate. % 0.001395675.
For the normal condition, task performance decreases to be 0.99\% as accurate. 
% For normal, that value is 0.0009894681.

% Why are people so sucky with gridlines when time 2 is smaller?
%  > maybe the gridline is making ppl double guess becuase instead of relying on the illusion herusitc (which would make it really easy - just choose time 1!, they are forced to actually compare line length as marked by the gridlines, which is tough and less precise.)
%  > relying on illusion is good in this case. gridline hurts (although overall they do better).
%  > What is it about the heuristics that ppl use that they are more accurate when f>h?

We computed participants' task accuracy based on whether the vertical distance between the two lines is bigger in Time 1 or Time 2. 
However, because we observed a low accuracy, 
%(e.g., task accuracy drops below chance as the ratio between Time 1 and Time 2 gets closer to 1), 
we suspect that participants were not actually making their decision by comparing the vertical distances.
% To better understand the driving mechanism behind our observations, we propose two perception models: perpendicular and equal triangles. 
We propose two alternative heuristics participants relied on when responding to the comparison task via two models:
\vspace{1mm}

\noindent \textbf{(1) Perpendicular:} this heuristic is about taking the perpendicular distance between the two lines, anchored on the bottom line at Time 1 and at Time 2, as shown in the middle panel in Figure~\ref{fig:fhmodeling}. This idea is inspired by existing work that suggests that sine illusion happens because people rely on the orthogonal instead of the vertical distance between the two lines~\cite{bu2020sinestream}.
\vspace{1mm}

\noindent \textbf{(2) Equal Triangle:} this heuristic takes the length of the line segment perpendicular to the angle bisector of the bottom and top lines, as shown in the right panel in Figure~\ref{fig:fhmodeling}. This is inspired by the `hull area' proxy from Jardine et al.~\cite{jardine2019perceptual} and the theory proposed by Day et al.~\cite{day1991sine} which suggest that participants could have considered the overall general dimensional area (i.e., similar to drawing a circle with radius equal to the delta between the two lines at the point of comparison) surrounding the points of comparison when making the decision. 
\vspace{1mm}

We re-compute the task accuracy by assuming that participants were relying on these two heuristics when comparing Time 1 and Time 2.
For example, for the perpendicular heuristic, we compute the perpendicular distance at Time 1 and Time 2. 
If the perpendicular distance at Time 1 is greater than Time 2, and participants selected Time 1 to be greater, even if the vertical distance at Time 1 is smaller, we would consider their response correct.
Under this setup, if the task accuracy increases, then we can infer that participants were more likely to rely on the perpendicular heuristic when completing the task rather than the vertical distance. 

Comparing the three line charts in Figure~\ref{fig:fhmodeling}, we see that participants' task accuracy increases under the assumptions of the perpendicular and equal triangle models. 
We conducted a t-test comparing the overall accuracy between these two models and found the equal triangle model to be the one with higher accuracy (t = $2.75$, p = $0.0060$). 

% To better understand the driving mechanism behind our observations, we propose two perception models: perpendicular and equal triangles. 
% data:  PERP and EQUAL
% t = -2.7505, df = 76337, p-value = 0.005953
% alternative hypothesis: true difference in means is not equal to 0
% 95 percent confidence interval:
%  -0.015792651 -0.002650186
% sample estimates:
% mean of x mean of y 
% 0.6836687 0.6928901 

% Both models compares the task accuracy assuming the participants used a different heuristic than the vertical difference. 
% We compute the task accuracy based on this different heuristic.
% If the task accuracy is higher, we can infer that this alternative heuristic more closely describes what participants actually compared in their task. 
% To answer this, we can turn to modeling what ppl actually compared.

% \vspace{2mm}
% \noindent \textbf{Comparing Perpendicular:} 
% perpendicular line vs. 

% \vspace{2mm}
% \noindent \textbf{Comparing Equal Triangle:} 
% equal angle line (line perpendicular to the angle bisector of the bottom and top lines)

\vspace{2mm}
\noindent \textbf{Takeaway:} When participants compare the deltas between two lines, they are less likely to rely on the vertical distance between the lines. Rather, they compute the orthogonal distance, best modeled by looking at the length of the line segment perpendicular to the angle bisector of the bottom and top lines.
% EQUAL has higher accuracy, is the better model.

%-------------------------------------------------------------------------
\subsection{When Time 1 is Actually Larger}
%We also examine how the three design conditions influenced response accuracy 
When the vertical distance between the two lines at Time 1 is \textit{larger} than that at Time 2, relying on the sine illusion would help the participants.
In these cases, participants performed \textit{worse} with the gridlines condition and \textit{best} with the dotted lines condition.  
When they viewed the dotted line charts, their accuracy was 4.18 times higher than the gridlines condition (p < $0.001$).
When they viewed the default normal condition, their accuracy was 2.75 times higher than the gridlines condition (p < $0.001$).
% dotted is 4.18 times better (p < $0.001$), normal is 2.75 times better (p < $0.001$).
However, despite this reversal in condition effectiveness, participants overall performed significantly better (t = $61.84$, p < $0.001$) in the scenario where Time 1 delta is actually larger (ratio of Time 1 over Time 2 is greater 1, $M_{accuracy}$ = $0.89$), compared to the scenario where Time 1 delta is smaller 
%the deltas between the two lines in Time 1 is smaller than Time 2 
(ratio of Time 1 over Time 2 is smaller than 1, $M_accuracy$ = $0.61$).
See annotation on the left-most panel in Figure~\ref{fig:fhmodeling}.

% Welch Two Sample t-test

% data:  fhratio_conditions_time1Larger$accuracy and fhratio_conditions_data$accuracy
% t = 61.839, df = 17975, p-value < 2.2e-16
% alternative hypothesis: true difference in means is not equal to 0
% 95 percent confidence interval:
%  0.2729367 0.2908054
% sample estimates:
% mean of x mean of y 
%  0.888656  0.606785 

We turn to the perpendicular and equal triangle models to better understand what might have driven this result. 
While further investigation is needed to draw causal conclusions, we suspect that the existence of the gridlines made participants double-guess their response instead of relying on the sine illusion, which decreased accuracy.
When the gridlines are present, participants can compare the length of the gridlines between the two lines at Time 1 and Time 2 to make their decisions. 
% When they engage in this length comparison task, they may be 
But length comparison is subjected to imprecision and perceptual bias, as length encoding is not an extremely precise encoding channel~\cite{mccoleman2021rethinking, ceja2020truth, cleveland1993model}.
As a result, they make mistakes and perform with lower accuracy.
However, when the gridlines are turned off, participants tend to rely on the sine illusion and compare the equal triangle distance for the task, which tends to result as Time 1 being perceived as larger.
Thus when Time 1 delta is actually larger, relying on the sine illusion, rather than making a length comparison, results in a higher overall accuracy across all conditions, and a relatively lower accuracy for gridlines. 
% TODO: ADD DOTS IN FIGURE 5 TO SHOW THE WHEN TIME 1 IS ACTUALLY HIGHER POSITION!
% in a way receive more precise information about the vertical distance between the two lines at Time 1 and Time 2. 

% Why are people so sucky with gridlines when time 2 is smaller?
%  > maybe the gridline is making ppl double guess becuase instead of relying on the illusion herusitc (which would make it really easy - just choose time 1!, they are forced to actually compare line length as marked by the gridlines, which is tough and less precise.)
%  > relying on illusion is good in this case. gridline hurts (although overall they do better).
%  > What is it about the heuristics that ppl use that they are more accurate when f>h?

% To answer this, we can turn to modeling what ppl actually compared.

%-------------------------------------------------------------------------
%-------------------------------------------------------------------------
%-------------------------------------------------------------------------
%-------------------------------------------------------------------------
\section{Design Guidelines and Summary of Findings}
We produced a model that predicts the likelihood and severity of the sine illusion in line charts based on the ratio of the deltas between the two points of comparison.
We found that, in general, accuracy drops below chance when the ratio falls above 0.7 for default line charts, and below 75\% when the ratio falls above 0.5.
This means that when participants are comparing the difference between two lines at two different time points, if that difference is less than 50\%, then people will struggle.
And if that difference is less than 30\%, people will struggle significantly. 

This threshold can be mitigated by adding gridlines, especially when the gridlines are aligned with the points of comparison.
Adding gridlines can shift the threshold, such that people will only start to struggle when the ratio between the two vertical distances when the difference is less than 20\%. 
While we have not tested to identify optimal design options for the gridlines, existing work by Bartram et al. identified adopting an alpha value between 0.1 and 0.45 for gridlines might be the most preferred and effective~\cite{bartram2010whisper}.

In general, people make mistakes in this task because they are 
%relying on the orthogonal distance between the two lines, rather than 
not comparing the vertical distance between the two lines.
They are comparing the length of the line segments perpendicular to the angle bisector of the bottom and top lines, as modeled by the equal triangle heuristics. 
% What they are actually comparing is best modeled by the equal triangle model in our setting (which is about taking the length of the line segment perpendicular to the angle bisector of the bottom and top lines). 
% Use gridlines to help. don't use dotted lines. 
% people make heuristic based on equal triangle
% whisper don't scream to find best gridlines transparency!

%-------------------------------------------------------------------------
%-------------------------------------------------------------------------
%-------------------------------------------------------------------------
%-------------------------------------------------------------------------
\section{Limitation and Future Directions}
\label{limitations}
We modeled the effect of the ratio between the vertical distance at the two comparison points.
As shown in the left panel in Figure~\ref{fig:key}, factors such as the line slopes at the two points of comparison, could also influence the severity of sine illusions.
A preliminary logistic regression model predicting comparison accuracy with the line slopes at the two points of comparison suggests these factors to be significant (details can be found in the supplementary materials at \url{https://osf.io/kq87n/}).
Most notably, the slope of the top line at Time 1 seems to have the strongest effect on comparison accuracy (OR = $2.26$, p < $0.001$). 
Future work could examine these effects to create a predictive model that takes data values of the lines as input, and computes the relative slopes and deltas between two lines, and outputs a likelihood of the viewer seeing sine illusions.  
This model could inform the development of a dynamic scaling tool to mitigate sine illusions, potentially with the help of gridlines. 
% with gridlines assist to combat the bias

% x-axis bias (from the mirror results)
% We only looked at f and h.
% How does the other varied componetns on the chart affect things?
% regression results!
Further, we only tested one design of the line charts to avoid a combinatorial explosion of conditions. 
\add{Future work could consider alternative colors, line thickness, and line types with different spacing to validate the generalizability of our findings.}
Moreover, considering that the amount of data can impact perception~\cite{mccoleman2020no, mccoleman2021rethinking, xiong2019examining}, %and the sine illusion can manifest  multiple classes of data, 
future work can test line charts with more lines to explore the effect of data set size.
% for higher ecological validity. 
% different line colors, thickness, number of lines, 

Finally, the current experiment only modeled two heuristics. Participants could engage in other strategies when completing the task. 
For example, people might instinctively perceive the pairs of lines as shapes, so that changing perspectives produced by eye movements would not distort their percept~\cite{ninio2014geometrical}.. 
This would mean that, instead of reading the values following rules of graphical interpretation, participants might be comparing the width and height (major/minor axes) that shape. 
% s the constraint of producing a stable shape despite the
% changing perspectives produced by eye movements~\cite{ninio2014geometrical}.
Future experiment should consider think-aloud protocols or offline studies to elicit those strategies, to potentially identify even better models of perception.

 \acknowledgments{%
    The authors wish to thank Judith Uchidiuno, as well as NSF awards IIS-2237585 and IIS-2311575 for support.
 }

\bibliographystyle{abbrv-doi-hyperref}

\bibliography{ref}

\end{document}